\documentclass[aip,reprint]{revtex4-1}
\pdfoutput=1
\usepackage[version=3]{mhchem} 
\usepackage{epsfig}
\usepackage{graphicx}  

\usepackage{dcolumn}   
\usepackage{bm}        
\usepackage{amssymb}   
\usepackage{amsmath}

\usepackage{url}
\usepackage{color}

\usepackage{epstopdf}
\begin{document}

\author{Chris~Healey}
\affiliation{Institute for Quantum Science and Technology, University of Calgary, Calgary, AB, T2N 1N4, Canada}
\affiliation{National Institute for Nanotechnology, 11421 Saskatchewan Dr.\ NW,  Edmonton, AB T6G 2M9, Canada}
\author{Hamidreza~Kaviani}
\affiliation{Institute for Quantum Science and Technology, University of Calgary, Calgary, AB, T2N 1N4, Canada}
\affiliation{National Institute for Nanotechnology, 11421 Saskatchewan Dr.\ NW,  Edmonton, AB T6G 2M9, Canada}
\author{Marcelo~Wu}
\affiliation{Institute for Quantum Science and Technology, University of Calgary, Calgary, AB, T2N 1N4, Canada}
\affiliation{National Institute for Nanotechnology, 11421 Saskatchewan Dr.\ NW,  Edmonton, AB T6G 2M9, Canada}
\author{Behzad~Khanaliloo}
\affiliation{Institute for Quantum Science and Technology, University of Calgary, Calgary, AB, T2N 1N4, Canada}
\affiliation{National Institute for Nanotechnology, 11421 Saskatchewan Dr.\ NW,  Edmonton, AB T6G 2M9, Canada}
\author{Matthew~Mitchell}
\affiliation{Institute for Quantum Science and Technology, University of Calgary, Calgary, AB, T2N 1N4, Canada}
\affiliation{National Institute for Nanotechnology, 11421 Saskatchewan Dr.\ NW,  Edmonton, AB T6G 2M9, Canada}
\author{Aaron~C.~Hryciw}
\affiliation{nanoFAB Facility, University of Alberta, Edmonton, Alberta T6G 2R3, Canada}
\author{Paul~E.~Barclay}
\email{pbarclay@ucalgary.ca}
\affiliation{Institute for Quantum Science and Technology, University of Calgary, Calgary, AB, T2N 1N4, Canada}
\affiliation{National Institute for Nanotechnology, 11421 Saskatchewan Dr.\ NW,  Edmonton, AB T6G 2M9, Canada}

\title{Design and experimental demonstration of optomechanical paddle nanocavities}

\begin{abstract}

We present the design, fabrication and initial characterization of a paddle nanocavity consisting of a suspended sub-picogram nanomechanical resonator optomechanically coupled to a photonic crystal nanocavity.  The optical and mechanical properties of the paddle nanocavity can be systematically designed and optimized, and key characteristics including mechanical frequency easily tailored. Measurements under ambient conditions of a silicon paddle nanocavity demonstrate an optical mode with quality factor~$Q_o\sim$~6000 near $1550$~nm, and optomechanical coupling to several mechanical resonances with frequencies  $\omega_m/2\pi\sim~12-64$~MHz, effective masses~$m_\text{eff}\sim~350-650$~fg, and mechanical quality factors~$Q_m\sim~44-327$.  Paddle nanocavities are promising for optomechanical sensing and nonlinear optomechanics experiments.

\end{abstract}

\maketitle

Ultrasensitive measurement and control of local dynamics on the nanoscale can be achieved with cavity optomechanical systems whose optical modes are coupled to mechanical resonances~\cite{ref:aspelmeyer2014co, ref:favero2014fom}. The interaction between photons and phonons within these devices can be  enhanced by optical nanocavities with  wavelength--scale dimensions~\cite{ref:eichenfield2009oc, ref:eichenfield2009apn}, and many recent theoretical proposals and experiments have shown that it is possible to optically probe the quantum properties of mesoscopic mechanical systems. Applications of cavity optomechanics \cite{ref:metcalfe2014aco} include ultrasensitive displacement and force detection~\cite{ref:arcizet2006hso, ref:li2009bat, ref:anetsberger2010mnm, ref:krause2012ahm, ref:liu2012wcs, ref:wu2014ddo}, optical cooling of a mechanical mode to its quantum ground-state~\cite{ref:chan2011lcn}, and optical squeezing~\cite{ref:safavi2013sls}.  Cavity optomechanical coupling can be both dispersive and dissipative \cite{ref:wu2014ddo}, and in some systems, including `membrane in the middle'  systems~\cite{ref:sankey2010stn, ref:flowers2012fcb, ref:karuza2013tlq}, whispering gallery mode devices~\cite{ref:doolin2014nos, ref:brawley2014nom} and photonic crystal optomechanical cavities~\cite{2015arXiv150507291P} can have nonlinear contributions. Nonlinear optomechanical coupling is predicted to enable observation of quantum non-demolition (QND) measurements of phonon number~\cite{ref:thompson2008sdc, ref:gangat2011pnq}, measurement of phonon shot noise~\cite{ref:clerk2010qmp}, and mechanical cooling and squeezing~\cite{ref:bhattacharya2008otc, ref:nunnenkamp2010csq, ref:biancofiore2011qdo}. Optomechanical paddle nanocavities are predicted to have large nonlinear optomechanical coupling \cite{Kaviani:15, 2015arXiv150507291P} owing to their fg-scale effective masses $m_\text{eff}$, relatively low [MHz] mechanical frequencies $\omega_m$, and correspondingly large zero point fluctuation amplitudes $x_\text{zpf}^2=\hbar/2m_\text{eff}\omega_m$. Here we present a procedure for the optical and mechanical design of an optomechanical paddle nanocavity and experimentally demonstrate its optomechanical coupling.



\begin{figure}[b]
\begin{center}
\epsfig{figure=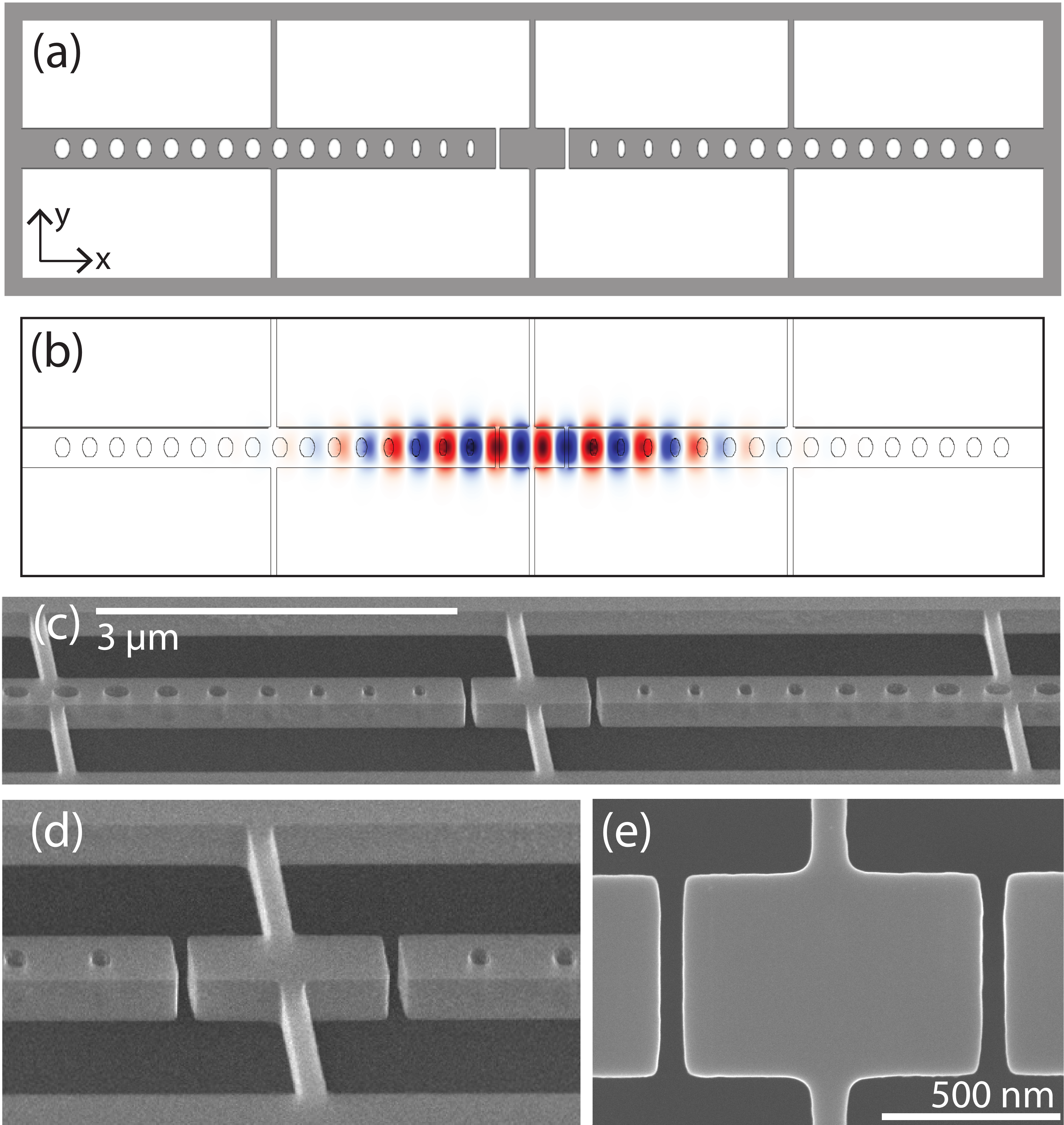, width=1\linewidth}
\caption{(a) Overview of the dielectric structure of a fabricable $\mathbf{m}$-sized paddle nanocavity. (b)~The FDTD-simulated electric field profile of the structure in~(a).  (c-e)~Scanning electron micrographs of a suspended paddle nanocavity nanofabricated in silicon.}
\label{introImage}
\end{center}
\end{figure}


The paddle nanocavity device demonstrated in this letter,  shown schematically and after fabrication in Figure~\ref{introImage}, consists of photonic crystal Bragg mirrors patterned in opposing nanocantilevers, with a low-frequency and small effective-mass  paddle mechanical oscillator suspended between them. When the nanobeam Bragg mirrors are patterned appropriately, the device forms a high quality factor ($Q_o$) optical cavity whose optical modes overlap with the mechanical resonances of the paddle.  Below we show how the device can be systematically designed and optimized to support high-$Q_o$ modes despite the large perturbation to the photonic crystal lattice created by the paddle. We then experimentally demonstrate optomechanical coupling between a high-$Q_o$ optical mode and both paddle and nanobeam mechanical resonances of a fabricated device.

\begin{figure}[t]
\begin{center}
\epsfig{figure=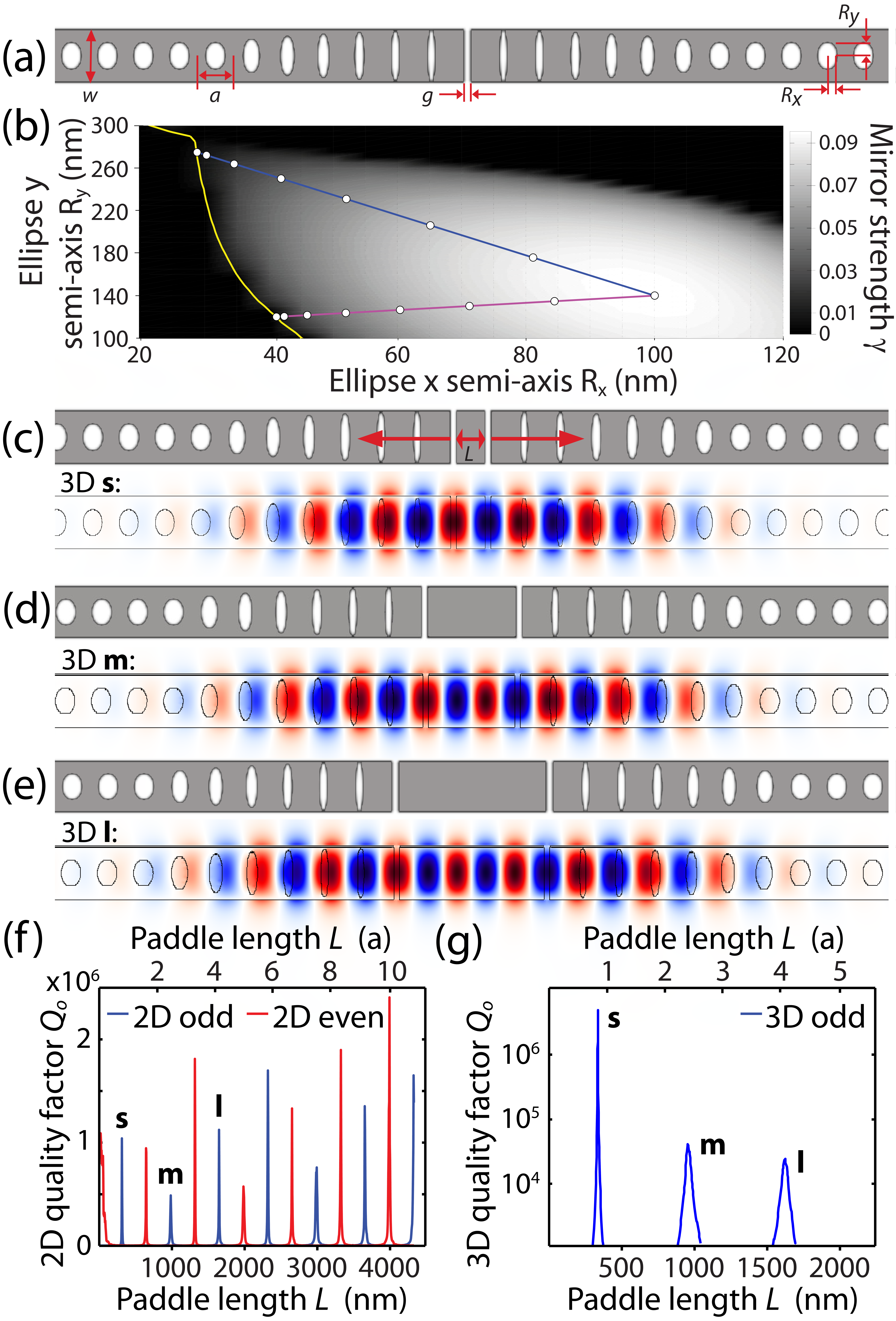, width=1\linewidth}
\caption{(a) Overview of a split-beam nanocavity.  (b)  Mirror strength $\gamma$ as a function of unit cell hole dimensions $R_{x}$ and $R_{y}$.  The blue upper (magenta lower) line shows the tapering trajectory for the optimal (fabricable) hole device design.  (c-e)~ Split-beam cavity from (a) with paddle length $L$ inserted in cavity volume, and simulated electric field profile of the  fundamental odd mode. $L$ set in (c-e) to values corresponding to the small $\mathbf{s}$, medium $\mathbf{m}$ and large $\mathbf{l}$ paddle sizes. The (f) 2D-FDTD simulated and (g) 3D-FDTD simulated odd (blue) and even-mode (red) $Q_o$ for varying $L$, showing oscillations at half-wavelength intervals.}
\label{design}
\end{center}
\end{figure}

The paddle nanocavities presented here use the split-beam photonic crystal nanocavity~\cite{Hryciw:13,ref:wu2014ddo}, developed by Hryciw et al., as a basis. This geometry is shown schematically in Fig.~\ref{design}(a), and will be described before analyzing the paddle nanocavity.  Split-beam structures can be deterministically designed\cite{Quan10,Quan11} to support a high-$Q_o$ even air-mode~\cite{Hryciw:13} and the structure as discussed here is intended to be fabricated from silicon-on-insulator (SOI) chips, with nanobeam width $w$ = 600 nm and thickness $t$ = 220 nm, lattice constant $a$ = 400 nm, and  hole semi-minor and semi-major axes $(R_{x},R_{y})$  that taper quadratically.  The tapering smoothly increases (decreases) the air filling fraction away from the nanocavity centre for an air mode (dielectric mode) whose optical intensity overlaps with the low-index air (high-index dielectric) regions of the structure.  Before introducing a gap in the nanocavity center, the design procedure determines  central hole dimensions that match the unit-cell band-edge with the target nanocavity resonance frequency~$\omega_o$, and the outer 'mirror' hole dimensions to maximize the mirror strength $\gamma=[(\omega_2-\omega_1)^2/(\omega_2+\omega_1)^2-(\omega_\textrm{o}-\omega_{mid})^2/(\omega_{mid})^2]^{1/2}$, where $\omega_1 (\omega_2)$ is the lower (upper) band edge frequency of the mirror region, and $\omega_{mid} = (\omega_2+\omega_1)/2$ is the mid-gap frequency.  Starting from an elliptical central hole with radii ($R_{x_\textrm{c}}$,~$R_{y_\textrm{c}}$)~=~(28.8,~275)~nm, we quadratically taper these dimensions over $N_c=7$ cavity holes with $R_{x_j,y_j}= R_{x_\textrm{c},y_\textrm{c}}+ (j/N_\textrm{c})^2(R_{x_\textrm{m},y_\textrm{m}}-R_{x_\textrm{c},y_\textrm{c}})$ for integer $j\in[-N_\textrm{c},N_\textrm{c}]$ to external mirror hole dimensions ($R_{x_\textrm{m}}$,~$R_{y_\textrm{m}}$)~=~(100,~140)~nm.  Figure~\ref{design}(b) shows the hole dimensions and the corresponding $\gamma$ for each hole in this design. A gap is introduced with width $g = 50\,\text{nm}$, determined by comparing the band structure of the gap and central hole unit cells~\cite{Hryciw:13}.  As previously found by Hryciw\cite{Hryciw:13}, a smoothly varying optical potential is achieved by matching the gap unit cell air-mode band-edge with the ideal central hole dielectric mode band-edge.  The resulting split-beam nanocavity is predicted by finite difference time domain (FDTD) simulations\cite{Hryciw:13,ref:oskooi2010mff} to  support an optical mode with $Q_o$~=~3.3$\times 10^{6}$ at a wavelength $\lambda_o$~=~1583~\text{nm}~($\omega_o/2\pi \sim 200$~THz). This high-$Q_o$ is in part due to the highly elliptical shape of the cavity holes which resemble the gap.


\begin{figure}[t]
\begin{center}
\epsfig{figure=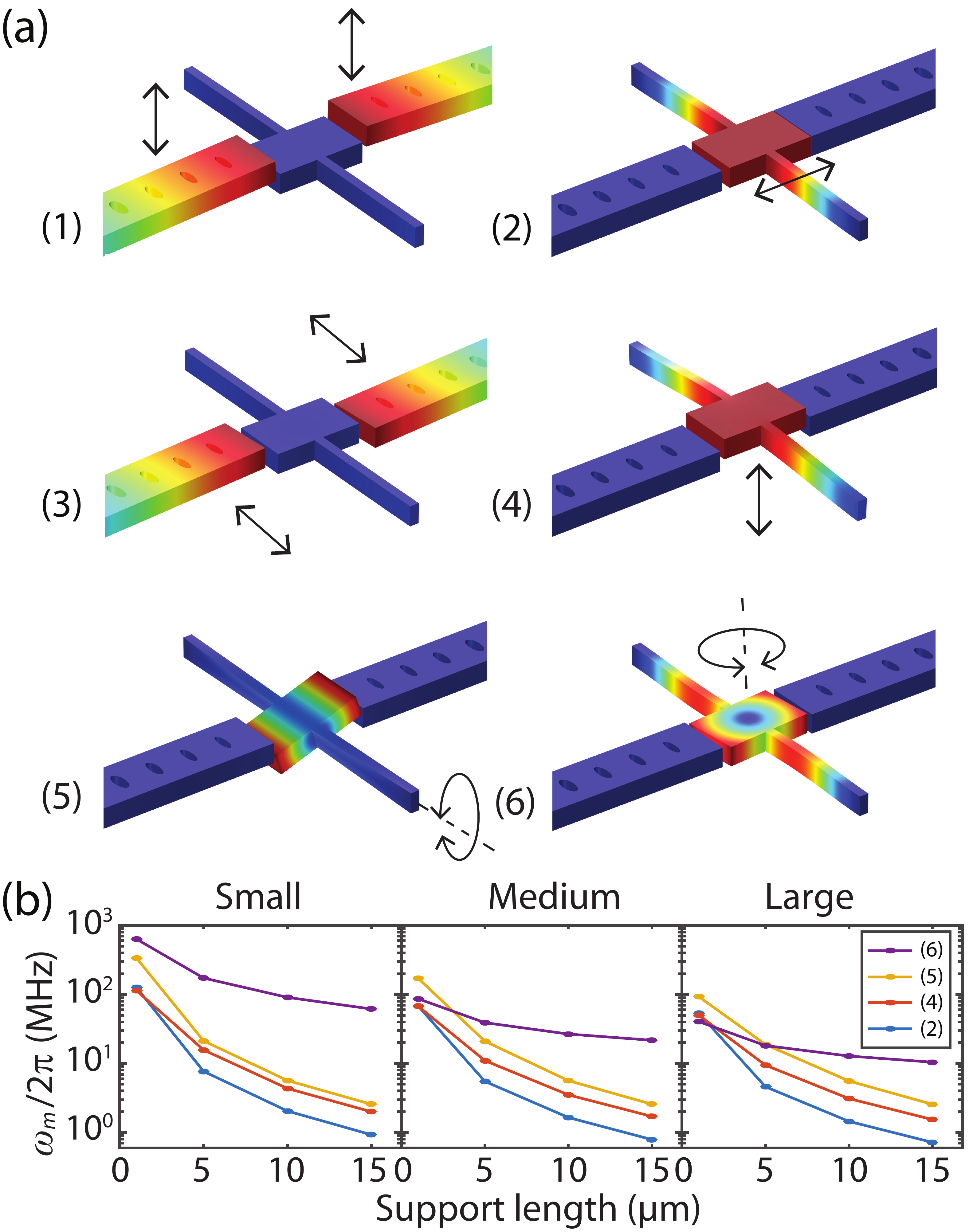, width=1\linewidth}
\caption{(a)  Displacement profiles of mechanical modes (1) - (6) as discussed in the text.  All motions are exaggerated.  (b)  The mechanical frequencies $\omega_m$ of paddle resonances as a function of support length for the three paddle sizes.}
\label{mechImage}
\end{center}
\end{figure}


To create a paddle nanocavity, the split-beam cantilevers are separated and an unpatterned dielectric block of length $L$ with the nominal waveguide cross-section is inserted between them, as shown in Fig.~\ref{design}(c).  As three-dimensional (3D) optical simulations can be time consuming~\cite{Quan10},  parameter searches in two-dimensions (2D) are first used to target optical modes with high-$Q_o$ within a chosen frequency range, followed by  3D simulations to optimize parameters.  We take advantage of the three-fold symmetry of the structure to reduce the computation time,  as we are interested in the lowest-frequency TE-like~($y$-odd, $z$-even) optical mode eigenfrequencies. To compensate for the lack of vertical confinement  in the 2D simulations,  a reduced effective index is used for silicon $n_\text{Si,eff} = 2.8 <n_\text{Si}$ such that the eigenfrequencies of 2D simulations roughly match 3D results.  Figure~\ref{design}(f) shows the results of 2D simulations of the even and odd $x$ symmetry modes for paddle length $L$ varying with high resolution ($\sim 10^3$ points).  $Q_o$ is found to oscillate as a function of $L$, with high-$Q_o$ values spaced in $L$ by integer wavelengths $\lambda_o=2an_\text{eff}$ for a given $x$ symmetry.  This is consistent with the result of Quan et al.\cite{Quan10} for an unpatterned waveguide capped by photonic crystal mirrors.  


In a realistic device, the paddle needs to be suspended by supports connected to the surrounding chip.  To minimize radiation loss introduced by the supports, they should be connected to the paddle at a node of the nanocavity field.  Hence, we only consider the odd optical modes of devices with supports connected to the paddle center.  As many quantum optical and optomechanical figures of merit scale with $Q_{o}/V$, where $V$ is the optical cavity mode volume defined by the peak field strength, we focus on the three smallest values of $L$  supporting high-$Q_o$ modes: $L$~=~334~nm, 1006~nm, and 1670~nm, which we label as small ($\mathbf{s}$), medium ($\mathbf{m}$) and large ($\mathbf{l}$) paddle lengths respectively.  Figure~\ref{design}(g) shows the predicted $Q_o(L)$ interpolated from approximately fifty 3D FDTD simulations in the neighbourhood of each targeted $L$, for a paddle nanocavity without supports.  We find high-$Q_{o}$ peaks at $L$ values in close agreement with predictions from the 2D simulations in Fig.~\ref{design}(f). For the $\mathbf{s}$, $\mathbf{m}$ and $\mathbf{l}$ paddle nanocavities, as shown in Fig.~\ref{design}(c),(d) and (e), we find 3D simulated $Q_{o}>$ 3.53$\times 10^{6}$, 3.43$\times 10^{4}$, and 1.72$\times 10^{4}$ for paddle lengths 334~nm, 964~nm, and 1604~nm respectively, all at $\lambda_o \sim$~1584~nm.  The corresponding $E_y$ electric field profiles for these nanocavities are plotted in Fig.~\ref{design}(c-e), and show that the optical modes are tightly confined within the tapered hole region and overlap with the paddle.  

The smallest nominal hole semi-minor axes of $\sim30$ nm in the design used above are challenging to fabricate.  Using the fabrication process discussed below, we can consistently realize holes with ($R_{x}$,~$R_{y}$)~$\sim$~(40,~100)~nm.  To design a `fabricable' device within this constraint, we designed a paddle nanocavity tapering from central hole dimensions of (41.2,~120)~nm, with the resulting hole dimensions and $\gamma$ shown in Fig.\ \ref{design}(b). As before, we replace the central hole with a paddle of length $L$ from the $\mathbf{m}$ optimized design above, separated from the cantilevers by 50 nm gaps. The resulting structure is shown schematically in Fig.~\ref{introImage}(a).   Without supports and before optimization, the simulated quality factor of this design is $Q_o \sim 1.48 \times 10^{4}$.  To realize a suspended paddle, we add 100~nm wide, 1.5~$\mu$m long centre supports. The resulting $Q_{o}$ increases to $2.34 \times 10^{4}$.  This indicates that for this design, scattering from the supports does not limit $Q_o$, and that $L$ from the ideal design is not optimal for the supported fabricable structure. After re-optimizing $L$, we find $Q_o$ to be within simulation uncertainty of $Q_o$ of the optimized device without supports. The simulated electric field profile of this device is shown in Fig.~\ref{introImage}(b). 


\begin{figure}[t]
\begin{center}
\epsfig{figure=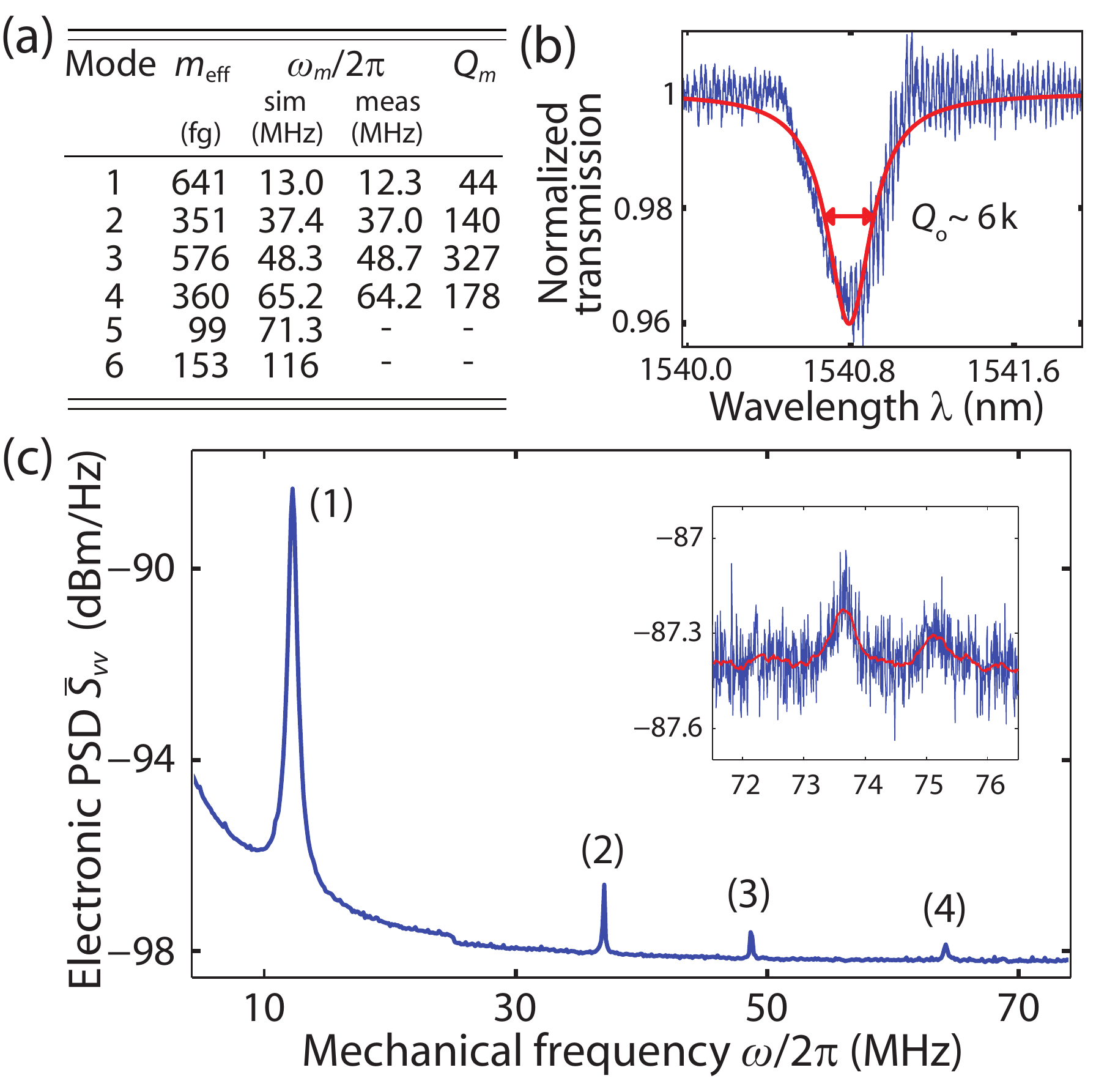, width=1\linewidth}
\caption{(a)  The mechanical mode effective masses $m_\text{eff}$, simulated (sim) and measured (meas) mechanical frequencies $\omega_{m}/2\pi$, and measured ambient quality factors $Q_m$.  (b) Fiber taper transmission for wavelengths scanned across nanocavity mode.  (c) Electronic power spectral density of the photodetected taper transmission with the laser source tuned within the nanocavity resonance. The mechanical modes are identified as discussed in the text.}
\label{results}
\end{center}
\end{figure}


The mechanical properties of the device, in particular $\omega_m$, are also affected by the paddle and support design.  Figure~\ref{mechImage}(a) shows six mechanical resonance displacement profiles, labeled~(1)-(6), calculated using a finite element simulation (COMSOL) of the $\mathbf{m}$ fabricable device.  Resonances (1) and (3) are characterized by the cantilevers moving  up and down, and side-to-side, respectively.  Resonances (2), (4), (5) and (6) involve the paddle moving in-plane along the device axis, up-and-down, torsionally, and rotationally, respectively.  We liken the axial motion of resonance (2) to the membrane-in-the-middle scheme, and note that the torsional resonance could be useful for sensing applications, for example in torque magnetometry~\cite{ref:losby2012tsc}. The resonance frequencies $\omega_{m}$ as a function of support length (100 nm support width) for the paddle modes are shown in Fig.~\ref{mechImage}(b), indicating that $\omega_m$ can be tuned by several orders of magnitude.  Figure \ref{results}(a) tabulates $m_\text{eff}$ for each mode. As a result of the wavelength-scale dimensions of the paddle, $m_\text{eff}$ are sub-pg, with the torsional mode exhibiting the smallest $m_\text{eff} \sim 99\,\text{fg}$.
   
To experimentally study paddle nanocavities, we fabricated devices from 220~nm thick silicon-on-oxide~(SOI).  SOI chips were coated with ZEP-520A resist, the design pattern was exposed with a 30~keV~Raith~150-Two electron beam lithography system, and was transferred to the Si-layer with a C$_4$F$_8$/SF$_6$ reactive ion etch.  The sacrificial 3~$\mu$m thick silicon oxide layer was selectively removed using hydrofluoric acid, creating suspended devices.  Figure~\ref{introImage}(c-e) shows SEM images of a typical device (type $\mathbf{m}$).  Note the measured gap width and paddle length in Fig.~\ref{introImage}(e) are approximately 55~nm and 898~nm respectively, which are within 10\% of the nominally designed values.

A dimpled, near-field fiber taper probe was used to test the paddle nanocavities~\cite{Hryciw:15}.  Light from a tunable diode laser (New Focus Velocity) at $\sim$1541~nm was input to the fiber taper evanescently coupled to the device. The fiber was positioned using 50 nm resolution stepper motor stages either hovering above the paddle, or contacting one of the cantilevers. All measurements were performed under ambient conditions.  The transmitted light was split using a 10:90 fiber coupler, with the outputs detected using low- (Newport 1623) and high-bandwidth (Newport 1811) detectors.   Detector signals were monitored with a data acquisition card and a real-time spectrum analyzer (Tektronix RSA 5106A) for investigating the optical mode and mechanical resonances, respectively.  The spectrum in Fig.~\ref{results}(b) shows the fiber taper transmission as a function of wavelength when the dimpled fiber is contacting a paddle nanocavity cantilever. A dip in transmission characteristic of evanescent coupling to the nanocavity optical mode is observed, with a linewidth corresponding to $Q_o\sim 6000$.  The electronic power spectral density~(PSD) $\bar{S}_{vv}(\omega)$ measured by the RSA when the laser  wavelength is fixed within this linewidth  is shown in Fig.~\ref{results}(c). Four peaks resulting from optomechanical transduction of the thermally-driven paddle resonances were clearly observed, with frequencies closely matching the simulated $\omega_m$ of the fabricated structure, as tabulated in Fig.~\ref{results}(a). Also tabulated are the measured mechanical quality factors, $Q_m$, of these resonances. For the measurements shown here, the low-$Q_m$ are a result of viscous damping from the ambient environment.   Two other low signal-to-noise mechanical resonances were also observed near 74~MHz, shown in the inset to Fig.~\ref{results}(c).  One peak is expected to be the torsional resonance (5), and the other a nonlinear harmonic of the axial sliding resonance~(2)~\cite{Kaviani:15, ref:doolin2014nos}. Although paddle nanocavities are completely symmetric in design and in theory have no intrinsic linear optomechanical coupling~\cite{Johnson02,Kaviani:15}, fabrication imperfections and fiber-induced  dissipative and dispersive optomechanical coupling~\cite{Hryciw:15} enable the observed optomechanical transduction. The nature of this coupling requires further measurements of the wavelength and fiber position dependence of the observed signal \cite{ref:wu2014ddo}.  

Enhancements to the optomechanical measurement sensitivity could be realized by operating under vacuum conditions to increase $Q_m$ by several orders of magnitude~\cite{PhysRevE.85.056313}. Combined with adjusting the support dimensions to lower $\omega_m$ to increase the thermal amplitude of the resonances, these improvements will allow higher signal-to-noise measurements, and may allow unambiguous discrimination of the torsional mode. These changes would also benefit characterization of nonlinear optomechanical coupling, whose signal strength scales with $\omega_{m}^{-4}$. This is of particular interest since analysis by Kaviani~et.~al~\cite{Kaviani:15} of $\mathbf{m}$ devices predict a quadratic optomechanical coupling coefficient $g^{(2)} > 2 \pi \times 400$ MHz/nm$^{2}$, and a single photon to two phonon optomechanical coupling rate $\Delta \omega_{0}>2 \pi \times 16$ Hz, well above the rates observed in similar systems~\cite{ref:lee2014mod, ref:doolin2014nos}.  We have also predicted  that the $\mathbf{s}$ and $\mathbf{l}$ designs have $g^{(2)}/2\pi$ of 100 and 550~MHz/nm$^2$, respectively.  Future nonlinear quantum optomechanics experiments, for example observation of phonon shot noise \cite{ref:clerk2010qmp}, are predicted to be possible using optimized paddle nanocavity devices \cite{Kaviani:15}. In addition to improving $Q_m$, reducing fabrication imperfections and increasing minimum features sizes to increase $Q_o$ are necessary to realize such experiments.

In conclusion, we have systematically designed and experimentally demonstrated a high-$Q_o$ paddle nanocavity, and observed optomechanical transduction of thermomechanical motion of several resonances of this device with $\sim 100-640$~fg effective mass. These devices are promising for future applications including torque magnetometry \cite{ref:losby2012tsc},  nanomechanical sensing \cite{ref:aspelmeyer2014co}, and nonlinear optomechanics. 

We thank the staff of the nanoFAB facility at the University of Alberta and at the National Institute for Nanotechnology for their technical support.  This work was funded by the National Research Council Canada (NRC), Natural Science and Engineering Research Council of Canada (NSERC), the Canada Foundation for Innovation (CFI) and Alberta Innovates Technology Futures (AITF).

\bibliography{nanoBib}

\begin{thebibliography}{35}%
\makeatletter
\providecommand \@ifxundefined [1]{%
 \@ifx{#1\undefined}
}%
\providecommand \@ifnum [1]{%
 \ifnum #1\expandafter \@firstoftwo
 \else \expandafter \@secondoftwo
 \fi
}%
\providecommand \@ifx [1]{%
 \ifx #1\expandafter \@firstoftwo
 \else \expandafter \@secondoftwo
 \fi
}%
\providecommand \natexlab [1]{#1}%
\providecommand \enquote  [1]{``#1''}%
\providecommand \bibnamefont  [1]{#1}%
\providecommand \bibfnamefont [1]{#1}%
\providecommand \citenamefont [1]{#1}%
\providecommand \href@noop [0]{\@secondoftwo}%
\providecommand \href [0]{\begingroup \@sanitize@url \@href}%
\providecommand \@href[1]{\@@startlink{#1}\@@href}%
\providecommand \@@href[1]{\endgroup#1\@@endlink}%
\providecommand \@sanitize@url [0]{\catcode `\\12\catcode `\$12\catcode
  `\&12\catcode `\#12\catcode `\^12\catcode `\_12\catcode `\%12\relax}%
\providecommand \@@startlink[1]{}%
\providecommand \@@endlink[0]{}%
\providecommand \url  [0]{\begingroup\@sanitize@url \@url }%
\providecommand \@url [1]{\endgroup\@href {#1}{\urlprefix }}%
\providecommand \urlprefix  [0]{URL }%
\providecommand \Eprint [0]{\href }%
\providecommand \doibase [0]{http://dx.doi.org/}%
\providecommand \selectlanguage [0]{\@gobble}%
\providecommand \bibinfo  [0]{\@secondoftwo}%
\providecommand \bibfield  [0]{\@secondoftwo}%
\providecommand \translation [1]{[#1]}%
\providecommand \BibitemOpen [0]{}%
\providecommand \bibitemStop [0]{}%
\providecommand \bibitemNoStop [0]{.\EOS\space}%
\providecommand \EOS [0]{\spacefactor3000\relax}%
\providecommand \BibitemShut  [1]{\csname bibitem#1\endcsname}%
\let\auto@bib@innerbib\@empty
\bibitem [{\citenamefont {Aspelmeyer}, \citenamefont {Kippenberg},\ and\
  \citenamefont {Marquardt}(2014)}]{ref:aspelmeyer2014co}%
  \BibitemOpen
  \bibfield  {author} {\bibinfo {author} {\bibfnamefont {M.}~\bibnamefont
  {Aspelmeyer}}, \bibinfo {author} {\bibfnamefont {T.~J.}\ \bibnamefont
  {Kippenberg}}, \ and\ \bibinfo {author} {\bibfnamefont {F.}~\bibnamefont
  {Marquardt}},\ }\href@noop {} {\bibfield  {journal} {\bibinfo  {journal}
  {Rev. Mod. Phys.}\ }\textbf {\bibinfo {volume} {86}},\ \bibinfo {pages}
  {1391} (\bibinfo {year} {2014})}\BibitemShut {NoStop}%
\bibitem [{\citenamefont {Favero}\ and\ \citenamefont
  {Marquardt}(2014)}]{ref:favero2014fom}%
  \BibitemOpen
  \bibfield  {author} {\bibinfo {author} {\bibfnamefont {I.}~\bibnamefont
  {Favero}}\ and\ \bibinfo {author} {\bibfnamefont {F.}~\bibnamefont
  {Marquardt}},\ }\href@noop {} {\bibfield  {journal} {\bibinfo  {journal} {New
  Journal of Physics}\ }\textbf {\bibinfo {volume} {16}},\ \bibinfo {pages}
  {085006} (\bibinfo {year} {2014})}\BibitemShut {NoStop}%
\bibitem [{\citenamefont {Eichenfield}\ \emph
  {et~al.}(2009{\natexlab{a}})\citenamefont {Eichenfield}, \citenamefont
  {Chan}, \citenamefont {Camacho}, \citenamefont {Vahala},\ and\ \citenamefont
  {Painter}}]{ref:eichenfield2009oc}%
  \BibitemOpen
  \bibfield  {author} {\bibinfo {author} {\bibfnamefont {M.}~\bibnamefont
  {Eichenfield}}, \bibinfo {author} {\bibfnamefont {J.}~\bibnamefont {Chan}},
  \bibinfo {author} {\bibfnamefont {R.}~\bibnamefont {Camacho}}, \bibinfo
  {author} {\bibfnamefont {K.}~\bibnamefont {Vahala}}, \ and\ \bibinfo {author}
  {\bibfnamefont {O.}~\bibnamefont {Painter}},\ }\href@noop {} {\bibfield
  {journal} {\bibinfo  {journal} {Nature}\ }\textbf {\bibinfo {volume} {462}},\
  \bibinfo {pages} {78} (\bibinfo {year} {2009}{\natexlab{a}})}\BibitemShut
  {NoStop}%
\bibitem [{\citenamefont {Eichenfield}\ \emph
  {et~al.}(2009{\natexlab{b}})\citenamefont {Eichenfield}, \citenamefont
  {Camacho}, \citenamefont {Chan}, \citenamefont {Vahala},\ and\ \citenamefont
  {Painter}}]{ref:eichenfield2009apn}%
  \BibitemOpen
  \bibfield  {author} {\bibinfo {author} {\bibfnamefont {M.}~\bibnamefont
  {Eichenfield}}, \bibinfo {author} {\bibfnamefont {R.}~\bibnamefont
  {Camacho}}, \bibinfo {author} {\bibfnamefont {J.}~\bibnamefont {Chan}},
  \bibinfo {author} {\bibfnamefont {K.~J.}\ \bibnamefont {Vahala}}, \ and\
  \bibinfo {author} {\bibfnamefont {O.}~\bibnamefont {Painter}},\ }\href@noop
  {} {\bibfield  {journal} {\bibinfo  {journal} {Nature}\ }\textbf {\bibinfo
  {volume} {459}},\ \bibinfo {pages} {550} (\bibinfo {year}
  {2009}{\natexlab{b}})}\BibitemShut {NoStop}%
\bibitem [{\citenamefont {Metcalfe}(2014)}]{ref:metcalfe2014aco}%
  \BibitemOpen
  \bibfield  {author} {\bibinfo {author} {\bibfnamefont {M.}~\bibnamefont
  {Metcalfe}},\ }\href@noop {} {\bibfield  {journal} {\bibinfo  {journal}
  {Applied Physics Reviews}\ }\textbf {\bibinfo {volume} {1}},\ \bibinfo {eid}
  {031105} (\bibinfo {year} {2014})}\BibitemShut {NoStop}%
\bibitem [{\citenamefont {Arcizet}\ \emph {et~al.}(2006)\citenamefont
  {Arcizet}, \citenamefont {Cohadon}, \citenamefont {Briant}, \citenamefont
  {Pinard}, \citenamefont {Heidmann}, \citenamefont {Mackowski}, \citenamefont
  {Michel}, \citenamefont {Pinard}, \citenamefont
  {Fran\ifmmode~\mbox{\c{c}}\else \c{c}\fi{}ais},\ and\ \citenamefont
  {Rousseau}}]{ref:arcizet2006hso}%
  \BibitemOpen
  \bibfield  {author} {\bibinfo {author} {\bibfnamefont {O.}~\bibnamefont
  {Arcizet}}, \bibinfo {author} {\bibfnamefont {P.-F.}\ \bibnamefont
  {Cohadon}}, \bibinfo {author} {\bibfnamefont {T.}~\bibnamefont {Briant}},
  \bibinfo {author} {\bibfnamefont {M.}~\bibnamefont {Pinard}}, \bibinfo
  {author} {\bibfnamefont {A.}~\bibnamefont {Heidmann}}, \bibinfo {author}
  {\bibfnamefont {J.-M.}\ \bibnamefont {Mackowski}}, \bibinfo {author}
  {\bibfnamefont {C.}~\bibnamefont {Michel}}, \bibinfo {author} {\bibfnamefont
  {L.}~\bibnamefont {Pinard}}, \bibinfo {author} {\bibfnamefont
  {O.}~\bibnamefont {Fran\ifmmode~\mbox{\c{c}}\else \c{c}\fi{}ais}}, \ and\
  \bibinfo {author} {\bibfnamefont {L.}~\bibnamefont {Rousseau}},\ }\href
  {\doibase 10.1103/PhysRevLett.97.133601} {\bibfield  {journal} {\bibinfo
  {journal} {Phys. Rev. Lett.}\ }\textbf {\bibinfo {volume} {97}},\ \bibinfo
  {pages} {133601} (\bibinfo {year} {2006})}\BibitemShut {NoStop}%
\bibitem [{\citenamefont {Li}, \citenamefont {Pernice},\ and\ \citenamefont
  {Tang}(2009)}]{ref:li2009bat}%
  \BibitemOpen
  \bibfield  {author} {\bibinfo {author} {\bibfnamefont {M.}~\bibnamefont
  {Li}}, \bibinfo {author} {\bibfnamefont {W.~H.~P.}\ \bibnamefont {Pernice}},
  \ and\ \bibinfo {author} {\bibfnamefont {H.~X.}\ \bibnamefont {Tang}},\
  }\href@noop {} {\bibfield  {journal} {\bibinfo  {journal} {Nat. Nano.}\
  }\textbf {\bibinfo {volume} {4}},\ \bibinfo {pages} {377} (\bibinfo {year}
  {2009})}\BibitemShut {NoStop}%
\bibitem [{\citenamefont {Anetsberger}\ \emph {et~al.}(2010)\citenamefont
  {Anetsberger}, \citenamefont {Gavartin}, \citenamefont {Arcizet},
  \citenamefont {Unterreithmeier}, \citenamefont {Weig}, \citenamefont
  {Gorodetsky}, \citenamefont {Kotthaus},\ and\ \citenamefont
  {Kippenberg}}]{ref:anetsberger2010mnm}%
  \BibitemOpen
  \bibfield  {author} {\bibinfo {author} {\bibfnamefont {G.}~\bibnamefont
  {Anetsberger}}, \bibinfo {author} {\bibfnamefont {E.}~\bibnamefont
  {Gavartin}}, \bibinfo {author} {\bibfnamefont {O.}~\bibnamefont {Arcizet}},
  \bibinfo {author} {\bibfnamefont {Q.~P.}\ \bibnamefont {Unterreithmeier}},
  \bibinfo {author} {\bibfnamefont {E.~M.}\ \bibnamefont {Weig}}, \bibinfo
  {author} {\bibfnamefont {M.~L.}\ \bibnamefont {Gorodetsky}}, \bibinfo
  {author} {\bibfnamefont {J.~P.}\ \bibnamefont {Kotthaus}}, \ and\ \bibinfo
  {author} {\bibfnamefont {T.~J.}\ \bibnamefont {Kippenberg}},\ }\href@noop {}
  {\bibfield  {journal} {\bibinfo  {journal} {Phys. Rev. A}\ }\textbf {\bibinfo
  {volume} {82}},\ \bibinfo {pages} {061804} (\bibinfo {year}
  {2010})}\BibitemShut {NoStop}%
\bibitem [{\citenamefont {Krause}\ \emph {et~al.}(2012)\citenamefont {Krause},
  \citenamefont {Winger}, \citenamefont {Blasius}, \citenamefont {Lin},\ and\
  \citenamefont {Painter}}]{ref:krause2012ahm}%
  \BibitemOpen
  \bibfield  {author} {\bibinfo {author} {\bibfnamefont {A.~G.}\ \bibnamefont
  {Krause}}, \bibinfo {author} {\bibfnamefont {M.}~\bibnamefont {Winger}},
  \bibinfo {author} {\bibfnamefont {T.~D.}\ \bibnamefont {Blasius}}, \bibinfo
  {author} {\bibfnamefont {W.}~\bibnamefont {Lin}}, \ and\ \bibinfo {author}
  {\bibfnamefont {O.}~\bibnamefont {Painter}},\ }\href@noop {} {\bibfield
  {journal} {\bibinfo  {journal} {Nat. Photon.}\ }\textbf {\bibinfo {volume}
  {6}},\ \bibinfo {pages} {768} (\bibinfo {year} {2012})}\BibitemShut {NoStop}%
\bibitem [{\citenamefont {Liu}\ \emph {et~al.}(2012)\citenamefont {Liu},
  \citenamefont {Miao}, \citenamefont {Aksyuk},\ and\ \citenamefont
  {Srinivasan}}]{ref:liu2012wcs}%
  \BibitemOpen
  \bibfield  {author} {\bibinfo {author} {\bibfnamefont {Y.}~\bibnamefont
  {Liu}}, \bibinfo {author} {\bibfnamefont {H.}~\bibnamefont {Miao}}, \bibinfo
  {author} {\bibfnamefont {V.}~\bibnamefont {Aksyuk}}, \ and\ \bibinfo {author}
  {\bibfnamefont {K.}~\bibnamefont {Srinivasan}},\ }\href@noop {} {\bibfield
  {journal} {\bibinfo  {journal} {Opt. Express}\ }\textbf {\bibinfo {volume}
  {20}},\ \bibinfo {pages} {18268} (\bibinfo {year} {2012})}\BibitemShut
  {NoStop}%
\bibitem [{\citenamefont {Wu}\ \emph {et~al.}(2014)\citenamefont {Wu},
  \citenamefont {Hryciw}, \citenamefont {Healey}, \citenamefont {Lake},
  \citenamefont {Jayakumar}, \citenamefont {Freeman}, \citenamefont {Davis},\
  and\ \citenamefont {Barclay}}]{ref:wu2014ddo}%
  \BibitemOpen
  \bibfield  {author} {\bibinfo {author} {\bibfnamefont {M.}~\bibnamefont
  {Wu}}, \bibinfo {author} {\bibfnamefont {A.~C.}\ \bibnamefont {Hryciw}},
  \bibinfo {author} {\bibfnamefont {C.}~\bibnamefont {Healey}}, \bibinfo
  {author} {\bibfnamefont {D.~P.}\ \bibnamefont {Lake}}, \bibinfo {author}
  {\bibfnamefont {H.}~\bibnamefont {Jayakumar}}, \bibinfo {author}
  {\bibfnamefont {M.~R.}\ \bibnamefont {Freeman}}, \bibinfo {author}
  {\bibfnamefont {J.~P.}\ \bibnamefont {Davis}}, \ and\ \bibinfo {author}
  {\bibfnamefont {P.~E.}\ \bibnamefont {Barclay}},\ }\href {\doibase
  10.1103/PhysRevX.4.021052} {\bibfield  {journal} {\bibinfo  {journal} {Phys.
  Rev. X}\ }\textbf {\bibinfo {volume} {4}},\ \bibinfo {pages} {021052}
  (\bibinfo {year} {2014})}\BibitemShut {NoStop}%
\bibitem [{\citenamefont {Chan}\ \emph {et~al.}(2011)\citenamefont {Chan},
  \citenamefont {Alegre}, \citenamefont {Safavi-Naeini}, \citenamefont {Hill},
  \citenamefont {Krause}, \citenamefont {Groblacher}, \citenamefont
  {Aspelmeyer},\ and\ \citenamefont {Painter}}]{ref:chan2011lcn}%
  \BibitemOpen
  \bibfield  {author} {\bibinfo {author} {\bibfnamefont {J.}~\bibnamefont
  {Chan}}, \bibinfo {author} {\bibfnamefont {T.~P.~M.}\ \bibnamefont {Alegre}},
  \bibinfo {author} {\bibfnamefont {A.~H.}\ \bibnamefont {Safavi-Naeini}},
  \bibinfo {author} {\bibfnamefont {J.~T.}\ \bibnamefont {Hill}}, \bibinfo
  {author} {\bibfnamefont {A.}~\bibnamefont {Krause}}, \bibinfo {author}
  {\bibfnamefont {S.}~\bibnamefont {Groblacher}}, \bibinfo {author}
  {\bibfnamefont {M.}~\bibnamefont {Aspelmeyer}}, \ and\ \bibinfo {author}
  {\bibfnamefont {O.}~\bibnamefont {Painter}},\ }\href@noop {} {\bibfield
  {journal} {\bibinfo  {journal} {Nature}\ }\textbf {\bibinfo {volume} {478}},\
  \bibinfo {pages} {89} (\bibinfo {year} {2011})}\BibitemShut {NoStop}%
\bibitem [{\citenamefont {Safavi-Naeini}\ \emph {et~al.}(2013)\citenamefont
  {Safavi-Naeini}, \citenamefont {Gr\"{o}blacher}, \citenamefont {Hill},
  \citenamefont {Chan}, \citenamefont {Aspelmeyer},\ and\ \citenamefont
  {Painter}}]{ref:safavi2013sls}%
  \BibitemOpen
  \bibfield  {author} {\bibinfo {author} {\bibfnamefont {A.~H.}\ \bibnamefont
  {Safavi-Naeini}}, \bibinfo {author} {\bibfnamefont {S.}~\bibnamefont
  {Gr\"{o}blacher}}, \bibinfo {author} {\bibfnamefont {J.~T.}\ \bibnamefont
  {Hill}}, \bibinfo {author} {\bibfnamefont {J.}~\bibnamefont {Chan}}, \bibinfo
  {author} {\bibfnamefont {M.}~\bibnamefont {Aspelmeyer}}, \ and\ \bibinfo
  {author} {\bibfnamefont {O.}~\bibnamefont {Painter}},\ }\href@noop {} {\
  \textbf {\bibinfo {volume} {500}},\ \bibinfo {pages} {185} (\bibinfo {year}
  {2013})}\BibitemShut {NoStop}%
\bibitem [{\citenamefont {Sankey}\ \emph {et~al.}(2010)\citenamefont {Sankey},
  \citenamefont {Yang}, \citenamefont {Zwickl}, \citenamefont {Jayich},\ and\
  \citenamefont {Harris}}]{ref:sankey2010stn}%
  \BibitemOpen
  \bibfield  {author} {\bibinfo {author} {\bibfnamefont {J.~C.}\ \bibnamefont
  {Sankey}}, \bibinfo {author} {\bibfnamefont {C.}~\bibnamefont {Yang}},
  \bibinfo {author} {\bibfnamefont {B.~M.}\ \bibnamefont {Zwickl}}, \bibinfo
  {author} {\bibfnamefont {A.~M.}\ \bibnamefont {Jayich}}, \ and\ \bibinfo
  {author} {\bibfnamefont {J.~G.~E.}\ \bibnamefont {Harris}},\ }\href@noop {}
  {\bibfield  {journal} {\bibinfo  {journal} {Nature Phys.}\ }\textbf {\bibinfo
  {volume} {6}},\ \bibinfo {pages} {707} (\bibinfo {year} {2010})}\BibitemShut
  {NoStop}%
\bibitem [{\citenamefont {Flowers-Jacobs}\ \emph {et~al.}(2012)\citenamefont
  {Flowers-Jacobs}, \citenamefont {Hoch}, \citenamefont {Sankey}, \citenamefont
  {Kashkanova}, \citenamefont {Jayich}, \citenamefont {Deutsch}, \citenamefont
  {Reichel},\ and\ \citenamefont {Harris}}]{ref:flowers2012fcb}%
  \BibitemOpen
  \bibfield  {author} {\bibinfo {author} {\bibfnamefont {N.}~\bibnamefont
  {Flowers-Jacobs}}, \bibinfo {author} {\bibfnamefont {S.}~\bibnamefont
  {Hoch}}, \bibinfo {author} {\bibfnamefont {J.}~\bibnamefont {Sankey}},
  \bibinfo {author} {\bibfnamefont {A.}~\bibnamefont {Kashkanova}}, \bibinfo
  {author} {\bibfnamefont {A.}~\bibnamefont {Jayich}}, \bibinfo {author}
  {\bibfnamefont {C.}~\bibnamefont {Deutsch}}, \bibinfo {author} {\bibfnamefont
  {J.}~\bibnamefont {Reichel}}, \ and\ \bibinfo {author} {\bibfnamefont
  {J.}~\bibnamefont {Harris}},\ }\href@noop {} {\bibfield  {journal} {\bibinfo
  {journal} {Applied Physics Letters}\ }\textbf {\bibinfo {volume} {101}},\
  \bibinfo {pages} {221109} (\bibinfo {year} {2012})}\BibitemShut {NoStop}%
\bibitem [{\citenamefont {Karuza}\ \emph {et~al.}(2013)\citenamefont {Karuza},
  \citenamefont {Galassi}, \citenamefont {Biancofiore}, \citenamefont
  {Molinelli}, \citenamefont {Natali}, \citenamefont {Tombesi}, \citenamefont
  {Di~Giuseppe},\ and\ \citenamefont {Vitali}}]{ref:karuza2013tlq}%
  \BibitemOpen
  \bibfield  {author} {\bibinfo {author} {\bibfnamefont {M.}~\bibnamefont
  {Karuza}}, \bibinfo {author} {\bibfnamefont {M.}~\bibnamefont {Galassi}},
  \bibinfo {author} {\bibfnamefont {C.}~\bibnamefont {Biancofiore}}, \bibinfo
  {author} {\bibfnamefont {C.}~\bibnamefont {Molinelli}}, \bibinfo {author}
  {\bibfnamefont {R.}~\bibnamefont {Natali}}, \bibinfo {author} {\bibfnamefont
  {P.}~\bibnamefont {Tombesi}}, \bibinfo {author} {\bibfnamefont
  {G.}~\bibnamefont {Di~Giuseppe}}, \ and\ \bibinfo {author} {\bibfnamefont
  {D.}~\bibnamefont {Vitali}},\ }\href@noop {} {\bibfield  {journal} {\bibinfo
  {journal} {Journal of Optics}\ }\textbf {\bibinfo {volume} {15}},\ \bibinfo
  {pages} {025704} (\bibinfo {year} {2013})}\BibitemShut {NoStop}%
\bibitem [{\citenamefont {Doolin}\ \emph {et~al.}(2014)\citenamefont {Doolin},
  \citenamefont {Hauer}, \citenamefont {Kim}, \citenamefont {MacDonald},
  \citenamefont {Ramp},\ and\ \citenamefont {Davis}}]{ref:doolin2014nos}%
  \BibitemOpen
  \bibfield  {author} {\bibinfo {author} {\bibfnamefont {C.}~\bibnamefont
  {Doolin}}, \bibinfo {author} {\bibfnamefont {B.}~\bibnamefont {Hauer}},
  \bibinfo {author} {\bibfnamefont {P.}~\bibnamefont {Kim}}, \bibinfo {author}
  {\bibfnamefont {A.}~\bibnamefont {MacDonald}}, \bibinfo {author}
  {\bibfnamefont {H.}~\bibnamefont {Ramp}}, \ and\ \bibinfo {author}
  {\bibfnamefont {J.}~\bibnamefont {Davis}},\ }\href@noop {} {\bibfield
  {journal} {\bibinfo  {journal} {Physical Review A}\ }\textbf {\bibinfo
  {volume} {89}},\ \bibinfo {pages} {053838} (\bibinfo {year}
  {2014})}\BibitemShut {NoStop}%
\bibitem [{\citenamefont {Brawley}\ \emph {et~al.}(2014)\citenamefont
  {Brawley}, \citenamefont {Vanner}, \citenamefont {Larsen}, \citenamefont
  {Schmid}, \citenamefont {Boisen},\ and\ \citenamefont
  {Bowen}}]{ref:brawley2014nom}%
  \BibitemOpen
  \bibfield  {author} {\bibinfo {author} {\bibfnamefont {G.}~\bibnamefont
  {Brawley}}, \bibinfo {author} {\bibfnamefont {M.}~\bibnamefont {Vanner}},
  \bibinfo {author} {\bibfnamefont {P.}~\bibnamefont {Larsen}}, \bibinfo
  {author} {\bibfnamefont {S.}~\bibnamefont {Schmid}}, \bibinfo {author}
  {\bibfnamefont {A.}~\bibnamefont {Boisen}}, \ and\ \bibinfo {author}
  {\bibfnamefont {W.}~\bibnamefont {Bowen}},\ }\href@noop {} {\bibfield
  {journal} {\bibinfo  {journal} {arXiv:1404.5746}\ } (\bibinfo {year}
  {2014})}\BibitemShut {NoStop}%
\bibitem [{\citenamefont {{Para{\"i}so}}\ \emph {et~al.}(2015)\citenamefont
  {{Para{\"i}so}}, \citenamefont {{Kalaee}}, \citenamefont {{Zang}},
  \citenamefont {{Pfeifer}}, \citenamefont {{Marquardt}},\ and\ \citenamefont
  {{Painter}}}]{2015arXiv150507291P}%
  \BibitemOpen
  \bibfield  {author} {\bibinfo {author} {\bibfnamefont {T.~K.}\ \bibnamefont
  {{Para{\"i}so}}}, \bibinfo {author} {\bibfnamefont {M.}~\bibnamefont
  {{Kalaee}}}, \bibinfo {author} {\bibfnamefont {L.}~\bibnamefont {{Zang}}},
  \bibinfo {author} {\bibfnamefont {H.}~\bibnamefont {{Pfeifer}}}, \bibinfo
  {author} {\bibfnamefont {F.}~\bibnamefont {{Marquardt}}}, \ and\ \bibinfo
  {author} {\bibfnamefont {O.}~\bibnamefont {{Painter}}},\ }\href@noop {}
  {\bibfield  {journal} {\bibinfo  {journal} {arXiv:1505.07291}\ } (\bibinfo
  {year} {2015})}\BibitemShut {NoStop}%
\bibitem [{\citenamefont {Thompson}\ \emph {et~al.}(2008)\citenamefont
  {Thompson}, \citenamefont {Zwickl}, \citenamefont {Jayich}, \citenamefont
  {Marquardt}, \citenamefont {Girvin},\ and\ \citenamefont
  {Harris}}]{ref:thompson2008sdc}%
  \BibitemOpen
  \bibfield  {author} {\bibinfo {author} {\bibfnamefont {J.~D.}\ \bibnamefont
  {Thompson}}, \bibinfo {author} {\bibfnamefont {B.~M.}\ \bibnamefont
  {Zwickl}}, \bibinfo {author} {\bibfnamefont {A.~M.}\ \bibnamefont {Jayich}},
  \bibinfo {author} {\bibfnamefont {F.}~\bibnamefont {Marquardt}}, \bibinfo
  {author} {\bibfnamefont {S.~M.}\ \bibnamefont {Girvin}}, \ and\ \bibinfo
  {author} {\bibfnamefont {J.~G.~E.}\ \bibnamefont {Harris}},\ }\href@noop {}
  {\bibfield  {journal} {\bibinfo  {journal} {Nature}\ }\textbf {\bibinfo
  {volume} {452}},\ \bibinfo {pages} {72} (\bibinfo {year} {2008})}\BibitemShut
  {NoStop}%
\bibitem [{\citenamefont {Gangat}, \citenamefont {Stace},\ and\ \citenamefont
  {Milburn}(2011)}]{ref:gangat2011pnq}%
  \BibitemOpen
  \bibfield  {author} {\bibinfo {author} {\bibfnamefont {A.~A.}\ \bibnamefont
  {Gangat}}, \bibinfo {author} {\bibfnamefont {T.~M.}\ \bibnamefont {Stace}}, \
  and\ \bibinfo {author} {\bibfnamefont {G.~J.}\ \bibnamefont {Milburn}},\
  }\href@noop {} {\bibfield  {journal} {\bibinfo  {journal} {New Journal of
  Physics}\ }\textbf {\bibinfo {volume} {13}},\ \bibinfo {pages} {043024}
  (\bibinfo {year} {2011})}\BibitemShut {NoStop}%
\bibitem [{\citenamefont {Clerk}, \citenamefont {Marquardt},\ and\
  \citenamefont {Harris}(2010)}]{ref:clerk2010qmp}%
  \BibitemOpen
  \bibfield  {author} {\bibinfo {author} {\bibfnamefont {A.~A.}\ \bibnamefont
  {Clerk}}, \bibinfo {author} {\bibfnamefont {F.}~\bibnamefont {Marquardt}}, \
  and\ \bibinfo {author} {\bibfnamefont {J.~G.~E.}\ \bibnamefont {Harris}},\
  }\href {\doibase 10.1103/PhysRevLett.104.213603} {\bibfield  {journal}
  {\bibinfo  {journal} {Phys. Rev. Lett.}\ }\textbf {\bibinfo {volume} {104}},\
  \bibinfo {pages} {213603} (\bibinfo {year} {2010})}\BibitemShut {NoStop}%
\bibitem [{\citenamefont {Bhattacharya}, \citenamefont {Uys},\ and\
  \citenamefont {Meystre}(2008)}]{ref:bhattacharya2008otc}%
  \BibitemOpen
  \bibfield  {author} {\bibinfo {author} {\bibfnamefont {M.}~\bibnamefont
  {Bhattacharya}}, \bibinfo {author} {\bibfnamefont {H.}~\bibnamefont {Uys}}, \
  and\ \bibinfo {author} {\bibfnamefont {P.}~\bibnamefont {Meystre}},\
  }\href@noop {} {\bibfield  {journal} {\bibinfo  {journal} {Physical Review
  A}\ }\textbf {\bibinfo {volume} {77}},\ \bibinfo {pages} {033819} (\bibinfo
  {year} {2008})}\BibitemShut {NoStop}%
\bibitem [{\citenamefont {Nunnenkamp}\ \emph {et~al.}(2010)\citenamefont
  {Nunnenkamp}, \citenamefont {B\o{}rkje}, \citenamefont {Harris},\ and\
  \citenamefont {Girvin}}]{ref:nunnenkamp2010csq}%
  \BibitemOpen
  \bibfield  {author} {\bibinfo {author} {\bibfnamefont {A.}~\bibnamefont
  {Nunnenkamp}}, \bibinfo {author} {\bibfnamefont {K.}~\bibnamefont
  {B\o{}rkje}}, \bibinfo {author} {\bibfnamefont {J.~G.~E.}\ \bibnamefont
  {Harris}}, \ and\ \bibinfo {author} {\bibfnamefont {S.~M.}\ \bibnamefont
  {Girvin}},\ }\href {\doibase 10.1103/PhysRevA.82.021806} {\bibfield
  {journal} {\bibinfo  {journal} {Phys. Rev. A}\ }\textbf {\bibinfo {volume}
  {82}},\ \bibinfo {pages} {021806} (\bibinfo {year} {2010})}\BibitemShut
  {NoStop}%
\bibitem [{\citenamefont {Biancofiore}\ \emph {et~al.}(2011)\citenamefont
  {Biancofiore}, \citenamefont {Karuza}, \citenamefont {Galassi}, \citenamefont
  {Natali}, \citenamefont {Tombesi}, \citenamefont {Di~Giuseppe},\ and\
  \citenamefont {Vitali}}]{ref:biancofiore2011qdo}%
  \BibitemOpen
  \bibfield  {author} {\bibinfo {author} {\bibfnamefont {C.}~\bibnamefont
  {Biancofiore}}, \bibinfo {author} {\bibfnamefont {M.}~\bibnamefont {Karuza}},
  \bibinfo {author} {\bibfnamefont {M.}~\bibnamefont {Galassi}}, \bibinfo
  {author} {\bibfnamefont {R.}~\bibnamefont {Natali}}, \bibinfo {author}
  {\bibfnamefont {P.}~\bibnamefont {Tombesi}}, \bibinfo {author} {\bibfnamefont
  {G.}~\bibnamefont {Di~Giuseppe}}, \ and\ \bibinfo {author} {\bibfnamefont
  {D.}~\bibnamefont {Vitali}},\ }\href@noop {} {\bibfield  {journal} {\bibinfo
  {journal} {Physical Review A}\ }\textbf {\bibinfo {volume} {84}},\ \bibinfo
  {pages} {033814} (\bibinfo {year} {2011})}\BibitemShut {NoStop}%
\bibitem [{\citenamefont {Kaviani}\ \emph {et~al.}(2015)\citenamefont
  {Kaviani}, \citenamefont {Healey}, \citenamefont {Wu}, \citenamefont
  {Ghobadi}, \citenamefont {Hryciw},\ and\ \citenamefont
  {Barclay}}]{Kaviani:15}%
  \BibitemOpen
  \bibfield  {author} {\bibinfo {author} {\bibfnamefont {H.}~\bibnamefont
  {Kaviani}}, \bibinfo {author} {\bibfnamefont {C.}~\bibnamefont {Healey}},
  \bibinfo {author} {\bibfnamefont {M.}~\bibnamefont {Wu}}, \bibinfo {author}
  {\bibfnamefont {R.}~\bibnamefont {Ghobadi}}, \bibinfo {author} {\bibfnamefont
  {A.}~\bibnamefont {Hryciw}}, \ and\ \bibinfo {author} {\bibfnamefont {P.~E.}\
  \bibnamefont {Barclay}},\ }\href@noop {} {\bibfield  {journal} {\bibinfo
  {journal} {Optica}\ }\textbf {\bibinfo {volume} {2}},\ \bibinfo {pages} {271}
  (\bibinfo {year} {2015})}\BibitemShut {NoStop}%
\bibitem [{\citenamefont {Hryciw}\ and\ \citenamefont
  {Barclay}(2013)}]{Hryciw:13}%
  \BibitemOpen
  \bibfield  {author} {\bibinfo {author} {\bibfnamefont {A.~C.}\ \bibnamefont
  {Hryciw}}\ and\ \bibinfo {author} {\bibfnamefont {P.~E.}\ \bibnamefont
  {Barclay}},\ }\href@noop {} {\bibfield  {journal} {\bibinfo  {journal} {Opt.
  Lett.}\ }\textbf {\bibinfo {volume} {38}},\ \bibinfo {pages} {1612} (\bibinfo
  {year} {2013})}\BibitemShut {NoStop}%
\bibitem [{\citenamefont {Quan}, \citenamefont {Deotare},\ and\ \citenamefont
  {Loncar}(2010)}]{Quan10}%
  \BibitemOpen
  \bibfield  {author} {\bibinfo {author} {\bibfnamefont {Q.}~\bibnamefont
  {Quan}}, \bibinfo {author} {\bibfnamefont {P.}~\bibnamefont {Deotare}}, \
  and\ \bibinfo {author} {\bibfnamefont {M.}~\bibnamefont {Loncar}},\
  }\href@noop {} {\bibfield  {journal} {\bibinfo  {journal} {App. Phys. Lett.}\
  }\textbf {\bibinfo {volume} {96}},\ \bibinfo {pages} {203102} (\bibinfo
  {year} {2010})}\BibitemShut {NoStop}%
\bibitem [{\citenamefont {Quan}\ and\ \citenamefont {Loncar}(2011)}]{Quan11}%
  \BibitemOpen
  \bibfield  {author} {\bibinfo {author} {\bibfnamefont {Q.}~\bibnamefont
  {Quan}}\ and\ \bibinfo {author} {\bibfnamefont {M.}~\bibnamefont {Loncar}},\
  }\href@noop {} {\bibfield  {journal} {\bibinfo  {journal} {Opt. Express}\
  }\textbf {\bibinfo {volume} {19}},\ \bibinfo {pages} {18529} (\bibinfo {year}
  {2011})}\BibitemShut {NoStop}%
\bibitem [{\citenamefont {Oskooi}\ \emph {et~al.}(2010)\citenamefont {Oskooi},
  \citenamefont {Roundy}, \citenamefont {Ibanescu}, \citenamefont {Bermel},
  \citenamefont {Joannopoulos},\ and\ \citenamefont
  {Johnson}}]{ref:oskooi2010mff}%
  \BibitemOpen
  \bibfield  {author} {\bibinfo {author} {\bibfnamefont {A.~F.}\ \bibnamefont
  {Oskooi}}, \bibinfo {author} {\bibfnamefont {D.}~\bibnamefont {Roundy}},
  \bibinfo {author} {\bibfnamefont {M.}~\bibnamefont {Ibanescu}}, \bibinfo
  {author} {\bibfnamefont {P.}~\bibnamefont {Bermel}}, \bibinfo {author}
  {\bibfnamefont {J.}~\bibnamefont {Joannopoulos}}, \ and\ \bibinfo {author}
  {\bibfnamefont {S.~G.}\ \bibnamefont {Johnson}},\ }\href@noop {} {\bibfield
  {journal} {\bibinfo  {journal} {Comp. Phys. Comm.}\ }\textbf {\bibinfo
  {volume} {181}},\ \bibinfo {pages} {687} (\bibinfo {year}
  {2010})}\BibitemShut {NoStop}%
\bibitem [{\citenamefont {Losby}\ \emph {et~al.}(2012)\citenamefont {Losby},
  \citenamefont {Burgess}, \citenamefont {Diao}, \citenamefont {Fortin},
  \citenamefont {Hiebert},\ and\ \citenamefont {Freeman}}]{ref:losby2012tsc}%
  \BibitemOpen
  \bibfield  {author} {\bibinfo {author} {\bibfnamefont {J.}~\bibnamefont
  {Losby}}, \bibinfo {author} {\bibfnamefont {J.}~\bibnamefont {Burgess}},
  \bibinfo {author} {\bibfnamefont {Z.}~\bibnamefont {Diao}}, \bibinfo {author}
  {\bibfnamefont {D.}~\bibnamefont {Fortin}}, \bibinfo {author} {\bibfnamefont
  {W.}~\bibnamefont {Hiebert}}, \ and\ \bibinfo {author} {\bibfnamefont
  {M.}~\bibnamefont {Freeman}},\ }\href@noop {} {\bibfield  {journal} {\bibinfo
   {journal} {J. Appl. Phys.}\ }\textbf {\bibinfo {volume} {111}},\ \bibinfo
  {pages} {07D305} (\bibinfo {year} {2012})}\BibitemShut {NoStop}%
\bibitem [{\citenamefont {Hryciw}\ \emph {et~al.}(2015)\citenamefont {Hryciw},
  \citenamefont {Wu}, \citenamefont {Khanaliloo},\ and\ \citenamefont
  {Barclay}}]{Hryciw:15}%
  \BibitemOpen
  \bibfield  {author} {\bibinfo {author} {\bibfnamefont {A.~C.}\ \bibnamefont
  {Hryciw}}, \bibinfo {author} {\bibfnamefont {M.}~\bibnamefont {Wu}}, \bibinfo
  {author} {\bibfnamefont {B.}~\bibnamefont {Khanaliloo}}, \ and\ \bibinfo
  {author} {\bibfnamefont {P.~E.}\ \bibnamefont {Barclay}},\ }\href {\doibase
  10.1364/OPTICA.2.000491} {\bibfield  {journal} {\bibinfo  {journal} {Optica}\
  }\textbf {\bibinfo {volume} {2}},\ \bibinfo {pages} {491} (\bibinfo {year}
  {2015})}\BibitemShut {NoStop}%
\bibitem [{\citenamefont {Johnson}\ \emph {et~al.}(2002)\citenamefont
  {Johnson}, \citenamefont {Ibanescu}, \citenamefont {Skorobogatiy},
  \citenamefont {Weisberg}, \citenamefont {Joannopoulos},\ and\ \citenamefont
  {Fink}}]{Johnson02}%
  \BibitemOpen
  \bibfield  {author} {\bibinfo {author} {\bibfnamefont {S.~G.}\ \bibnamefont
  {Johnson}}, \bibinfo {author} {\bibfnamefont {M.}~\bibnamefont {Ibanescu}},
  \bibinfo {author} {\bibfnamefont {M.~A.}\ \bibnamefont {Skorobogatiy}},
  \bibinfo {author} {\bibfnamefont {O.}~\bibnamefont {Weisberg}}, \bibinfo
  {author} {\bibfnamefont {J.~D.}\ \bibnamefont {Joannopoulos}}, \ and\
  \bibinfo {author} {\bibfnamefont {Y.}~\bibnamefont {Fink}},\ }\href@noop {}
  {\bibfield  {journal} {\bibinfo  {journal} {Phys. Rev. E}\ }\textbf {\bibinfo
  {volume} {65}},\ \bibinfo {pages} {066611} (\bibinfo {year}
  {2002})}\BibitemShut {NoStop}%
\bibitem [{\citenamefont {Svitelskiy}\ \emph {et~al.}(2012)\citenamefont
  {Svitelskiy}, \citenamefont {Sauer}, \citenamefont {Vick}, \citenamefont
  {Cheng}, \citenamefont {Liu}, \citenamefont {Freeman},\ and\ \citenamefont
  {Hiebert}}]{PhysRevE.85.056313}%
  \BibitemOpen
  \bibfield  {author} {\bibinfo {author} {\bibfnamefont {O.}~\bibnamefont
  {Svitelskiy}}, \bibinfo {author} {\bibfnamefont {V.}~\bibnamefont {Sauer}},
  \bibinfo {author} {\bibfnamefont {D.}~\bibnamefont {Vick}}, \bibinfo {author}
  {\bibfnamefont {K.-M.}\ \bibnamefont {Cheng}}, \bibinfo {author}
  {\bibfnamefont {N.}~\bibnamefont {Liu}}, \bibinfo {author} {\bibfnamefont
  {M.~R.}\ \bibnamefont {Freeman}}, \ and\ \bibinfo {author} {\bibfnamefont
  {W.~K.}\ \bibnamefont {Hiebert}},\ }\href@noop {} {\bibfield  {journal}
  {\bibinfo  {journal} {Phys. Rev. E}\ }\textbf {\bibinfo {volume} {85}},\
  \bibinfo {pages} {056313} (\bibinfo {year} {2012})}\BibitemShut {NoStop}%
\bibitem [{\citenamefont {Lee}\ \emph {et~al.}(2014)\citenamefont {Lee},
  \citenamefont {Underwood}, \citenamefont {Mason}, \citenamefont {Shkarin},
  \citenamefont {Hoch},\ and\ \citenamefont {Harris}}]{ref:lee2014mod}%
  \BibitemOpen
  \bibfield  {author} {\bibinfo {author} {\bibfnamefont {D.}~\bibnamefont
  {Lee}}, \bibinfo {author} {\bibfnamefont {M.}~\bibnamefont {Underwood}},
  \bibinfo {author} {\bibfnamefont {D.}~\bibnamefont {Mason}}, \bibinfo
  {author} {\bibfnamefont {A.}~\bibnamefont {Shkarin}}, \bibinfo {author}
  {\bibfnamefont {S.}~\bibnamefont {Hoch}}, \ and\ \bibinfo {author}
  {\bibfnamefont {J.}~\bibnamefont {Harris}},\ }\href@noop {} {\bibfield
  {journal} {\bibinfo  {journal} {arXiv:1401.2968}\ } (\bibinfo {year}
  {2014})}\BibitemShut {NoStop}%
\end{thebibliography}%

\newpage
\newpage
\clearpage

\end{document}